\documentclass[preprint,12pt,5p,times,twocolumn]{elsarticle}

\usepackage{amssymb}
\usepackage{amsmath}

\newcounter{bla}

\journal{Computer Physics Communications}

\begin{document}

\begin{frontmatter}



\title{MsSpec-DFM (Dielectric function module): Towards a multiple scattering approach to plasmon description}


\author[a,b]{Aditi Mandal}
\author[b]{Sylvain Tricot}
\author[a]{Rakesh Choubisa}
\author[b]{Didier S\'{e}billeau\corref{author}}

\cortext[author] {Corresponding author.\\\textit{E-mail address:} didier.sebilleau@univ-rennes1.fr}
\address[a]{Department of Physics, Birla Institute of Technology and Science-Pilani, Pilani Campus, Pilani,  Rajasthan, 333031, India}
\address[b]{Univ. Rennes, CNRS, IPR (Institut de Physique de Rennes) - UMR 6251, F-35000 Rennes, France}

\begin{abstract}
We present here the MsSpec Dielectric Function module (MsSpec-DFM), which generates dielectric functions in an electron gas or a liquid, either isolated or embedded into an environment.

In addition to standard models such as the plasmon pole and the RPA, this module also provides more involved methods incorporating local field corrections (in order to account for correlations), Boltzmann-Vlasov hydrodynamical methods, the relaxation-damped Mermin and the diffusion-damped Hu-O'Connell methods, 
as well as moment-based methods using either a Nevanlinna function or a memory function.

Ultimately, through the use of form factors, the MsSpec-DFM module will be able to address a wide range of materials such as metals, semiconductors, including inversion layers, hetero-structures, superconductors, quantum wells, quantum wires, quantum dots, Dirac materials such as graphene, and liquids.


%
%
%

\end{abstract}


\end{frontmatter}



{\bf PROGRAM SUMMARY/NEW VERSION PROGRAM SUMMARY}

\begin{small}
\noindent
{\em Program Title:} MsSpec-DFM                                         \\
{\em CPC Library link to program files:} (to be added by Technical Editor) \\
{\em Developer's repository link:} (if available) \\
{\em Code Ocean capsule:} (to be added by Technical Editor)\\
{\em Licensing provisions(please choose one):} GPLv3 \\
{\em Programming language:}  FORTRAN 90                      \\
{\em Supplementary material:}                                 \\
{\em Journal reference of previous version:}*  None                \\
{\em Does the new version supersede the previous version?:}* Not applicable   \\
{\em Reasons for the new version:*}Not applicable \\
{\em Summary of revisions:}*Not applicable \\
{\em Nature of problem(approx. 50-250 words):}\\
{\em Solution method(approx. 50-250 words):}\\
{\em Additional comments including restrictions and unusual features (approx. 50-250 words):}\\
   \\

* Items marked with an asterisk are only required for new versions
of programs previously published in the CPC Program Library.\\
\end{small}

\section{Introduction}
\label{}
The MsSpec-DFM module, where DFM stands for Dielectric Function Module, is a new independent module of the MsSpec program package \cite{a1}. The core of MsSpec-DFM  consists in the calculation of model dielectric functions within the homogeneous electron gas or Fermi liquid, as well as in one-component or two-component plasmas, and classical liquids. From the purely multiple scattering point of view, it provides a flexible way to calculate the fluctuation potential that enters the quasi-boson modelling of the plasmon field pioneered by Hedin, Fujikawa and coworkers \cite{a2,a3}. But the possibilities of this module go much beyond the particular problem of the embedding of plasmons within the multiple scattering framework.
Indeed, providing a wide range of methods to compute dielectric functions in all sorts of materials,
it can also be used to calculate optical properties, cross-section of specific spectroscopies (Electron
Energy Loss Spectroscopy – EELS, X-ray inelastic scattering, Raman scattering , $\cdots$) or the
stopping power of particles of technological and of medical interest\footnote{Optical properties, cross-sections and stopping power will be added gradually in the package}. 
The scattering of electrons by atoms, molecules and solids is of extensive importance in many areas of physics, ranging from radiation physics and material analysis to Auger-electron spectroscopy (AES), and x-ray photoelectron spectroscopy (XPS). We refer to \cite{a1} and references therein for an in-
depth discussion of the various areas of application and developement of MsSpec for various spectroscopies. MsSpec-1.7, the latest version of the multiple scattering package basically computes the cluster, the potential and the cross-section of various spectroscopies namely photoelectron diffraction (PED), Auger electron diffraction (AED), X-ray absorption (XAS), low-energy
electron diffraction (LEED) and Auger photoelectron coincidence spectroscopy (APECS). For the cross-section, in order to cover a range of energies as wide as possible, several different algorithms
are provided to cover the energy range up to 1.5 keV. Multiple Scattering (MS) within the scattering path operator formalism allows to describe a whole
range of spectroscopies, with a minor cost for any of them once the MS part has been computed \cite{a4}. The 
present module is an addition to the MsSpec package that will utimately allow to describe PhotoEmission Energy Loss Spectroscopy (PEELS) \cite{a14}. This spectroscopy relies on the study and monitoring (as a function of the photoelectron exit angle, for example) of the plasmon peak.  This plasmon peak contains information on the bulk or surface dielectric function, depending on the type of plasmon peak considered. Therefore, an accurate modelling of the dielectric function is mandatory in order to describe properly PEELS. The ultimate goal of PEELS is to allow the extraction of the information on the dielectric function contained in the plasmon peak.   

For this, the MsSpec-DFM module provides a wide range of methods to compute the dielectric function beyond the famed RPA model which is limited by the fact that it does not account for electron correlations, it neglects  plasmon damping and it does not conserve basic quantities such as the number of particle, the momentum and the energy. In the present version of the module the non-homogeneous case is not treated. The module uses a constant density for the  material considered and it has  neither a crystal structure nor an electronic band structure. Phonon effects will be treated later in  phenomenological way by adding a phonon dielectric function. We are currently assessing the effect of the 
band structure on the dielectric function by performing 
ab initio dielectric function calculations with the Questaal code \cite{a31}.

\section{Theoretical Background}
Plasmon dispersion in the electron gas model has been widely studied theoretically as well as experimentally for last six decades. It has been an interesting topic in many-body phenomena and many theoretical calculations have been done which are based on various phenomenological models for the plasmon \cite{a5}. There has been considerable experimental and theoretical studies of both bulk and
surface-plasmon satellites of the x-ray photoemission spectra (XPS) of metals. In a typical core-level X-ray photoemission spectra in addition to a main sharp band there is plasmon loss bands with contribution mainly from two processes i.e. intrinsic and extrinsic processes.
The extrinsic losses were first discussed by Berglund and Spicer in their three step model \cite{a6} and since
then these semi-classical approaches has been widely used for practical purposes \cite{a7,a8,a9,a10}. Hedin et. al \cite{a2}
have compared the quantum one-step calculations with the semi-classical ones and used an electron-boson
Hamiltonian, where both the photoelectron and the core electron are coupled to bosonic-type excitations in the solid via standard fluctuation potentials. Later Fujikawa et. al \cite{a11} derived a general formula to describe overall features in core-level photoemission including plasmon losses and peak asymmetry due to X-ray singularity along with a quantum Landau formula which fully considers elastic scatterings before and after the losses. In connection with this, the main aspect behind this work is mainly the study of  different fluctuation potentials for the possible evaluation of self-energy and the role of fluctuation
potential generator from modeled dielectric functions to study various Dirac materials, Schrodinger materials and semiconductors.

\subsection{Fluctuation potential and Dielectric function}
Fluctuation potential can be termed as a coupling function arising between the electron and the quasibosons in a semi-infinite jellium model. It is the parameter which arises due to charge density fluctuation in the electronic system. 


This approach is fully quantum mechanical, so the recoil of photoelectron and its damping are considered. We do not dive here into details and discuss the widespread approximations for the fluctuation potentials. The fluctuation potentials available in the literature are the following: a)Plasmon pole, b)Inglesfield, c) Bechstedt. Beyond the available fluctuation potentials as stated earlier, the different fluctuation potential extracted from the dielectric functions using different local and dynamical local field corrections such as  RPA and improvements thereafter. Among which we will explore and explain UTIC1 (Utsumi-Ichimaru) type of fluctuation potential so that we can asses the role of fluctuation potential via dielectric function in the electron gas theory especially in photoemission.

The different models of fluctuation potentials have been obtained within linear response theory. As a consequence of the semi-infinite electron gas theory the fluctuation potentials can usually be written in the form:
\begin{equation}
V^{\mathbf{q}}(\mathbf{r})= \textit{f}(V_c(\mathbf{q}),\omega_{\mathbf{q}})e^{i\mathbf{q}.\mathbf{r}},
\end{equation}
where $V_c(\mathbf{q})$ is the Fourier transform of the Coulomb potential and $\omega_{\mathbf{q}} $ is the frequency of a plasmon of potential energy $\hbar\omega_{p} $ and momentum $\hbar\mathbf{q}$. As a consequence, the momentum vector of a plasmon will
be given by $\mathbf{q}=\mathbf{k}_{in}-\mathbf{k}_{sc} $ where $ \mathbf{k}_{in}$ is the momentum of the electron before the plasmon loss and $\mathbf{k}_{sc}$ is the the momentum of the electron after the plasmon loss. Further, simplifying the fluctuation potential in Eqn. (4) it can also be re-written in the form \cite{a11}
\begin{equation}
V^{\mathbf{q}}(\mathbf{r})=e^{i\mathbf{q}.\mathbf{r}}V^{\mathbf{q}}(z)
\end{equation}

Due to the axial symmetry with respect to the surface normal, the notation used here $\mathbf{r}=(\mathbf{r}_{\|},z) $ and $\mathbf{q}=(\mathbf{q}_{\|},q_{z}) $.

In the interest to treat more diverse type of materials and spectroscopies involving more energetic electrons, we tried with an alternative method to the problem of the description of the fluctuation potential. For this purpose, we use the definition that was given by Hedin and coworkers \cite{a2} where the fluctuation potential is defined as 
\begin{equation}
V^{\mathbf{q}}(\mathbf{r})=\left|\frac{V_{C}(\mathbf{q})}{\displaystyle{\frac{\partial\varepsilon(\mathbf{q},\omega)}{\partial\omega}\vert_{\omega=\omega(\mathbf{q})}}}\right|^{1/2}e^{i\mathbf{q}.\mathbf{r}}
\end{equation}

where $V_{C}(\mathbf{q})$ is the Fourier transform of the Coulomb potential and $\varepsilon(\mathbf{q},\omega)$ is the dielectric function of the system. The derivation is taken along the plasmon dispersion $\omega(\mathbf{q})$ which can be obtained by solving $\Re[\varepsilon(\mathbf{q},\omega)]=0$, where $\Re[ \quad]$ indicates the real part.

The second aspect is based on the fact that the information embedded into plasmon peaks has been studied less so far in photoemission. The photo-electron diffraction-like features surfaced back in 1990 through Osterwlader's group where they have exhibited the core-level peak for this feature \cite{a12}. It was further explored out in the seminal work of David and Godet work, where they have used these plasmon peaks to extract information of the system dielectric function from their energy contribution \cite{a13,a14}. This phenomena was coined as Photoemission Electron Energy Loss Spectroscopy (PEELS). Other works by Guzzo and coworkers \cite{a15,a16}, explores the other aspects of plasmon features which originated from valence band electrons. So we can infer that plasmon structures seems to be a promising tool for the extraction of information from spectroscopies and in our case from photoemission.

The dielectric function of a system essentially gives details of the response of the system to an external perturbation. It is determined by the properties of the system and its overall reciprocal action with the perturbing object. Several related quantities can be found in the literature, together with their association with various spectroscopies. For instance, it is well established that the cross-section of the EELS is related to the loss function.

The basic ingredients of dielectric function modeling 

\begin{equation}\label{}
\varepsilon(\mathbf{q},\omega)= \varepsilon_{1}(\mathbf{q},\omega)+ i\varepsilon_{2}(\mathbf{q},\omega)
\end{equation}

So, the dielectric function can be decomposed into its real and imaginary parts. Briefly, this knowledge of dielectric function explains electron-energy-loss experiments and also allows one to calculate the spectrum of collective excitations.

Then, the loss function $L(\mathbf{q}, \omega)$ is associated to dielectric function of the solid through following relation,

\begin{equation}\label{eq5}
L(\mathbf{q}, \omega) \, = \Im \left[\frac{-1}{ \varepsilon (\mathbf{q}, \omega)}\right] \, = \, \, \frac{\varepsilon_{2}(\mathbf{q}, \omega)}{\mid\varepsilon(\mathbf{q}, \omega)\mid^{2}}
\end{equation}

Similarly, for the case of 3D systems at T=0 K, the dynamical structure factor 
can be expressed as \cite{a17}

\begin{equation}\label{eq6}
S(\mathbf{q}, \omega) = \frac{\hbar}{\pi} \;  \frac{1}{\bar{n}} \;  \frac{1}{V_C(\mathbf{q})} \; 
\Im \left[\frac{-1}{\varepsilon(\mathbf{q}, \omega)}\right]
\end{equation}

As before, $V_C(\mathbf{q})$ is the Fourier transform of the Coulomb potential and $\bar{n}$ the constant electron density.
It describes the overall spectrum of excitations in the system as a function of the momentum transfer 
$\mathbf{q}$ and the energy transfer.

Likewise, the susceptibility, or density-density response function,  can be defined as

\begin{equation}\label{eq16}
\chi(\mathbf{q}, \omega) = \dfrac{1}{V_{C}(q)}\left[\dfrac{1}{\varepsilon(\mathbf{q}, \omega)}-1 \right]
\end{equation}

With these tools, we can access many different ways to model the dielectric function, and compute cross-sections.

Local field corrections are introduced through the local field corrected-RPA dielectric function

\begin{equation}
\varepsilon^{LF}(\vec{\bf q}, \omega) \, = \, 1 - \frac{V_C(\vec{\bf q}) \Pi^{RPA}(\vec{\bf q},\omega)}
{ 1 + V_C(\vec{\bf q}) G(\vec{\bf q},\omega) \Pi^{RPA}(\vec{\bf q},\omega)}
\end{equation}

where $G(\vec{\bf q},\omega)$ is the dynamical local field correction that introduces correlation effects into the otherwise \textit{non interacting} RPA model and $\Pi^{RPA}(\vec{\bf q},\omega)$ is the RPA polarization.

Random phase approximation (RPA) is essentially approximation for correlation function and an early application of the RPA was Lindhard's calculation of the dielectric function of the electron gas \cite{a18}. This is based on the elucidation of the de-localized electrons system as a homogeneous and non-interacting electron gas.

RPA alone has a certain number of constraints like it fails to conserve the number of particles, it does not contain correlation effects (they have to be added externally through local field corrections) and it is insufficient to incorporate plasmon damping outside the Landau regime. 

Hence, for solving this the Lindhard dielectric function was extended by Mermin \cite{a19} in the relaxation-time approximation, essentially where the collisions relax the electronic density matrix to a local equilibrium density matrix rather than to its uniform equilibrium value. This has also been incorporated in this module. The Mermin \cite{a20} dielectric function being defined as,

\begin{equation}
\varepsilon(\mathbf{q},\omega) \, = \, 1 + \frac{\displaystyle{\left( 1 + \frac{i}{\omega \tau} \right) }\;
\left[ \varepsilon^0(\mathbf{q},\omega + i/\tau) - 1 \right]}{1 + \displaystyle{\frac{i}{\omega \tau} \;
\left[ \frac{\varepsilon^0(\mathbf{q},\omega + i/ \tau) - 1}{\varepsilon^0(\mathbf{q},0) - 1} \right]}}
\end{equation}

where $\varepsilon^0(\mathbf{q},\omega)$ is the RPA dielectric function and $\bar{\omega} = \omega + i/\tau$ is a complex frequency incorporating damping through the relaxation time $\tau$.

Following the similar idea, Hu and O’Connell \cite{a21} included fluctuation effects arising from electron-electron and electron-impurity interactions, while generalizing the Lindhard dielectric function. As an example, they also studied Friedel oscillations and observed that these oscillations are damped due to the inclusion of fluctuation effects.

At present in this module, for 3D systems along with RPA there is a long list of various types of static local field corrections to choose from, such as the Hubbard model (HUBB), which takes only into account exchange
effect, the Pathak-Vashista correction (PVHF) \cite{a22}, and the Utsumi-Ichimaru (UTI1) \cite{a23} etc. and additionally relaxation times and temperature can be added as per the requirement of the system which is being studied for generating the respective dielectric function (or) the desired parameter file which can be opted from the output file generation segment.

\subsection{Nevanlinna and memory function methods}

An alternative approach of obtaining dielectric functions through a reconstruction from the first few moments was explored~\cite{a36}. It is well established that the dielectric function in RPA is found not to satisfy the compressibility sum rule and frequency moment sum rules \cite{a24}. Hence, these two independent approaches, the Nevanlinna approach \cite{a25} and the memory function approach \cite{a26} were studied and incorporated to explain the system in more realistic scenario. The main advantage of these two approaches is that conservation of the number of particles and correlation are built-in. 

Nevanlinna method is mathematically involved and this approach is based on the moments of the loss function essentially developed in the context of strongly coupled plasma. The main point is that it involves a reconstruction of the dielectric function in terms of characteristic frequencies $\omega_n$ that are functions of the first moments of the loss
function. 
We refer the interested readers to the review article by Tkachenko \cite{a27} for more detailed calculations. 



The memory function method was pioneered by Zwanzig \cite{a28} and Mori \cite{a29} who built this framework upon Kubo’s non-equilibrium statistical physics. The memory function method includes the effect of the fast variables into the dynamics of the slow variables. In order to connect this, we worked with time-correlation functions and response functions.

The connection to the dielectric function is made through the dynamical structure factor and the density-density response function and the relations are well described in Eqn.\ref{eq6} and \ref{eq16}. We observed that these two methods do improve upon the correlation-augmented RPA methods or the damping methods (Mermin, Hu-O’Connell).



\subsection{The electron-electron relaxation time (3D)}





In order to compute Mermin type dielectric functions, several electron-electron relaxation times are available, in addition to the possibility to provide it externally. A comparison of the different methods to compute this 
relaxation time is shown in the Fig\ref{fig4}.

Several relaxation schemes have been implemented in the MsSpec-DFM code such as the relaxation time used by Al'tshuler and Aronov ({\tt EE\_TYPE} = {\tt ALAR} and {\tt ALA2})\cite{a38},  Farm-Storz-Tom-Bokor ({\tt FSTB}) \cite{a39}. The Pines-Nozi\`{e}res relaxation time ({\tt PIN1}), valid in the high-density limit\cite{a40} and The Qian-Vignale lifetime of a quasi-particle in low density ({\tt QIVI}) and high density limit\cite{a41} are also available. Fig.\ref{fig4} compares these relaxation time methods.

\begin{figure}[htp]
\includegraphics[width = 1.0\columnwidth]{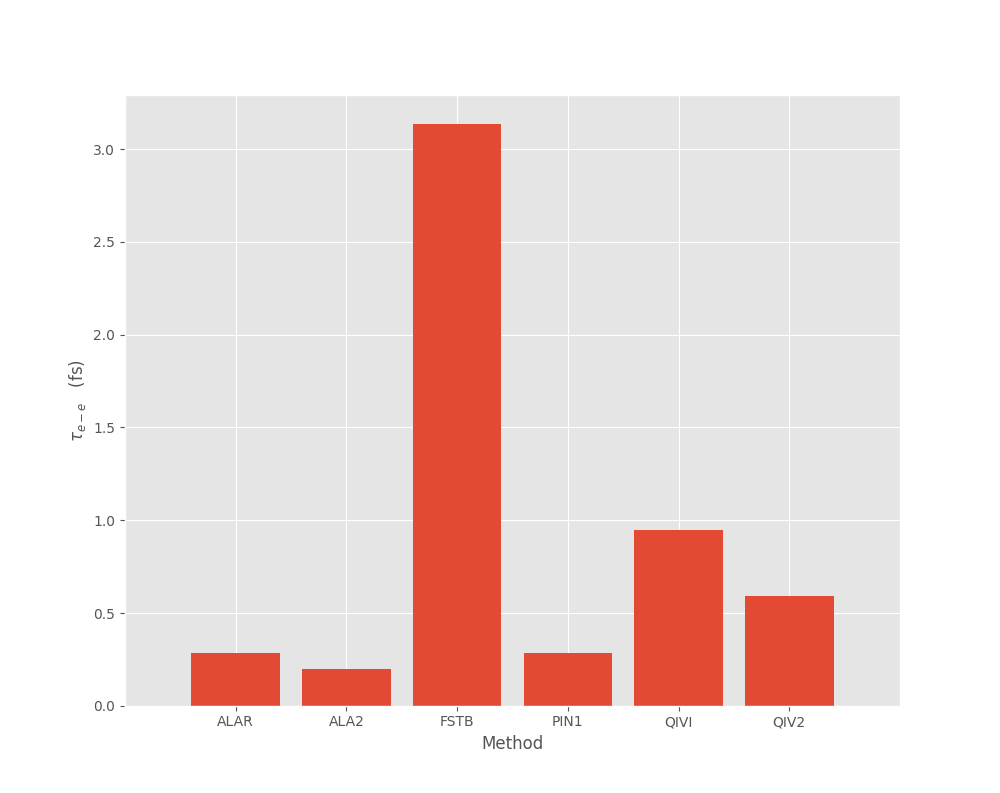}
\caption{Comparison of different $\tau_{e-e}$ relaxation times as calculated by MsSpec-DFM code}\label{fig4}
\end{figure}

\subsection{Dynamic structure factor $S(\vec{\bf q},\omega)$ in 3D}

\begin{figure}[htp]
\includegraphics[width = 1.0\columnwidth]{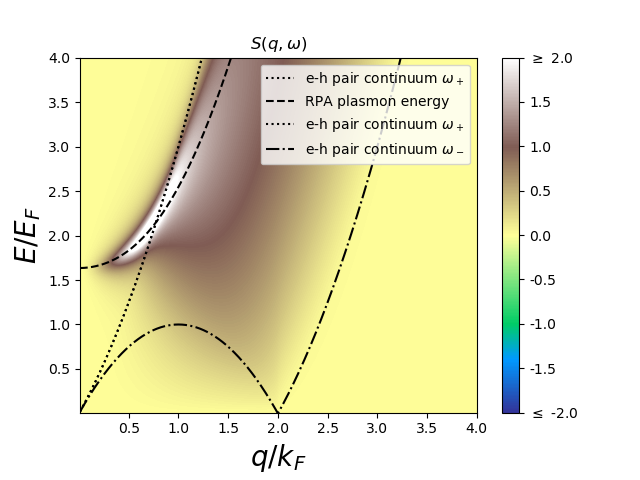}
\caption{Dynamic structure factor in the Mermin case for Ag}\label{fig5}
\end{figure}

Dynamic structure factors $S(\vec{\bf q},\omega)$ can also be computed by the code. An example based on the 
Mermin dielectric function is given in the Fig.\ref{fig5} for the case of Ag, with the choice of a relaxation time $\tau = 0.5$ fs. We note that this value of $\tau$ agrees with most of the $\tau_{e-e}$ displayed in figure \ref{fig4}.

\begin{figure}[htp]
\includegraphics[width = 1.0\columnwidth]{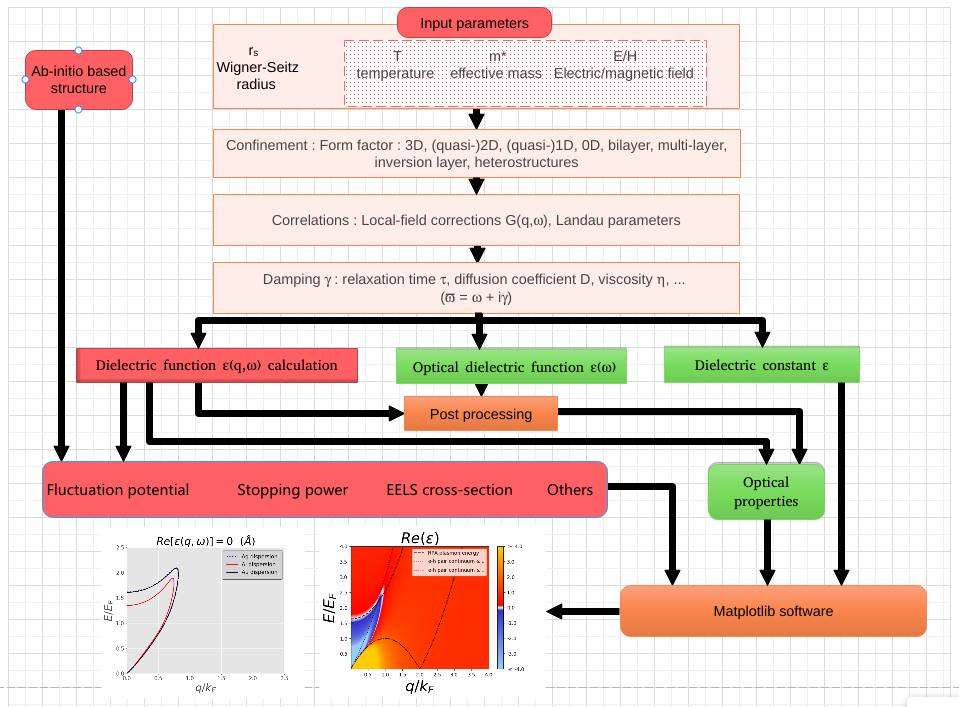}
\caption{Code flowchart}\label{fig2}
\end{figure}

\section{The code}
The code has been structured to deal with the many types of materials, external fields and temperature. The present version is restricted to 3D and 2D metals within the electron gas approach. Fermi liquid methods and other types of materials will be incorporated in the next upgrade, as well as external field dependence. Post processing functions such as the fluctuation potential, the stopping power, EELS/infrared cross-section are already under development for next release. Flow chart of the code in the Fig.\ref{fig2}.
The running of the code follows the four following steps:
\begin{enumerate}
    \item Editing and documenting with the input data the {\tt epsilon.dat} file (This file is self-documented)
    \item Compiling the code: {\tt make}
    \item Running the code: {\tt .\textbackslash eps}
    \item Plot the results: {\tt python3 ./graphEps.py re -2 2 4 4}
\end{enumerate}
In the plotting command the first two numbers refer to the lower and upper value of the z-component and the last two value are respectively the upper value of the x $(q/k_{F})$ and y $(E/E_{F})$ axis. The parameter after name of the python script should be {\tt re} to plot the real part and {\tt im} to plot the imaginary part. 

Several other plotting scripts are available to plot other quantities such as $S(\vec{\bf q},\omega)$ including one that subtracts two graphs to see the differences.

Note that the Landau continuum and theoretical RPA dispersion can be incorporated into the graph simply by transferring the {\tt elec\_hole.dat} and {\tt plas\_disp.dat} into the {\sl Plotting} directory.


\section{Test results}

We propose here two test cases: the RPA dielectric fucntion in Ag in 3D and 2D.

\subsection{Ag 3D}

In the {\tt epsilon.dat} file: 
In the general parameter section, we set {\tt Q\_MIN} = {\tt 0.010},
{\tt Q\_MAX} = {\tt 4.000} and {\tt N\_Q} = {\tt 1000} to vary $q / k_{F}$ 
from 0 to 4 (for {\tt Q\_MIN} = {\tt 0.000}, the calculation will diverge)
In the material properties section, we set {\tt RS} = {\tt 3.020} for Wigner-Seitz radius of Ag,
{\tt DIM} = {\tt 3D} in system's dimension part. In the dielectric function section, 
we set {\tt ESTDY} = {\tt DYNAMIC} and {\tt EPS\_T} = {\tt LONG} with {\tt D\_FUNC} = {\tt RPA1}.

\subsection{Ag 2D}

In the {\tt epsilon.dat} file: 
In the general parameter section, we set {\tt Q\_MIN} = {\tt 0.010},
{\tt Q\_MAX} = {\tt 4.000} and {\tt N\_Q} = {\tt 1000} to vary $q / k_{F}$ 
from 0 to 4 (for {\tt Q\_MIN} = {\tt 0.000}, the calculation will diverge)
In the material properties section, we set {\tt RS} = {\tt 3.020} for Wigner-Seitz radius of Ag,
{\tt DIM} = {\tt 2D} in system's dimension part. In the dielectric function section, 
we set {\tt ESTDY} = {\tt DYNAMIC} and {\tt EPS\_T} = {\tt LONG} with {\tt D\_FUNC} = {\tt RPA1}.

Finally in order to print the dielectric function into the {\tt diel\_func.dat} file, the {\tt I\_DF} switch in the output calculation {\sl Printing} section to be set to 1.

The only change with respect to the previous case is the dimension of the system from 3D to 2D.

\begin{figure}[htp]
\includegraphics[width = 1.0\columnwidth]{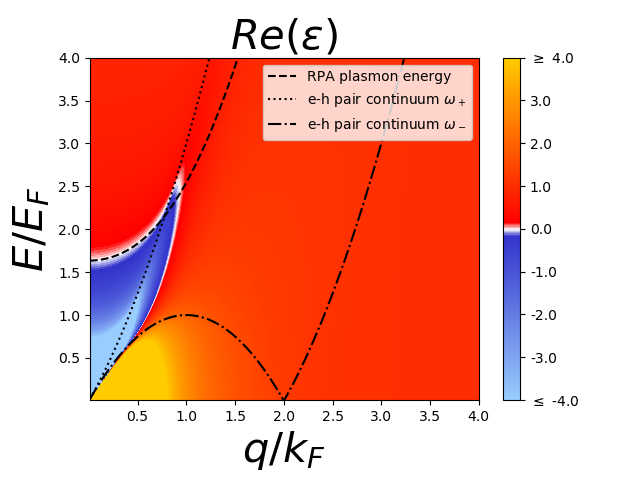}
\includegraphics[width = 1.0\columnwidth]{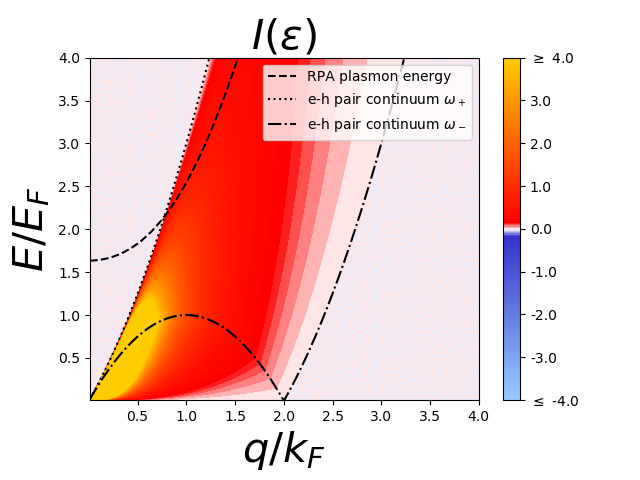}
\caption{3D RPA $\epsilon(\vec{\bf q},\omega)$ for Ag}\label{fig3}
\end{figure}

Our two test results are displayed in Fig.\ref{fig3} and Fig.\ref{fig4}, with the real part of the dielectric function at the top and the imaginary part at the bottom.
The Landau continuum is the region comprised between the two parabola $\omega_{+}$ and $\omega_{-}$. This is the region where excitation of single electron-hole pairs is the main loss mechanism. The Landau damping induces the decay of the plasmon into a single electron-hole pair. In Fig.\ref{fig3} and Fig.\ref{fig4} we observe there is no damping along the plasmon dispersion outside the Landau continuum. One way to overcome this problem and have a physically motivated damping outside the Landau continuum is to use the relaxation time-damped dielectric function, such as the Mermin ({\tt D\_FUNC} = {\tt MER2}), a diffusion-damped one ({\tt D\_FUNC} = {\tt HUCO}), or moment reconstructed dielectric function following the Nevanlinna method ({\tt D\_FUNC} = {\tt NEV3}) or using the memory function approach ({\tt D\_FUNC} = {\tt MEM3}) with a suitable choice of the corresponding relaxation functions.

\begin{figure}[htp]
\includegraphics[width = 1.0\columnwidth]{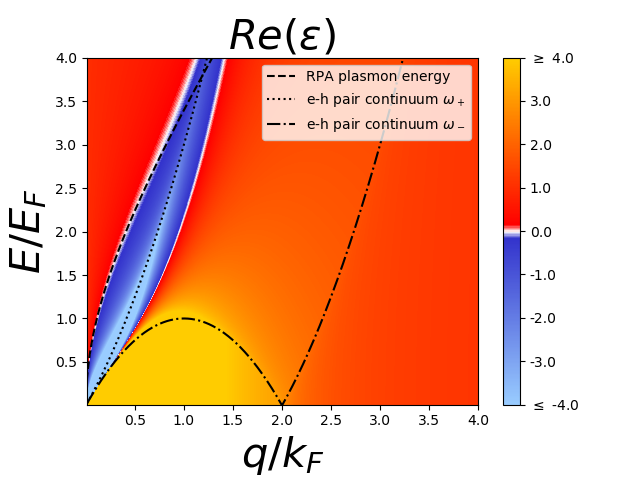}
\includegraphics[width = 1.0\columnwidth]{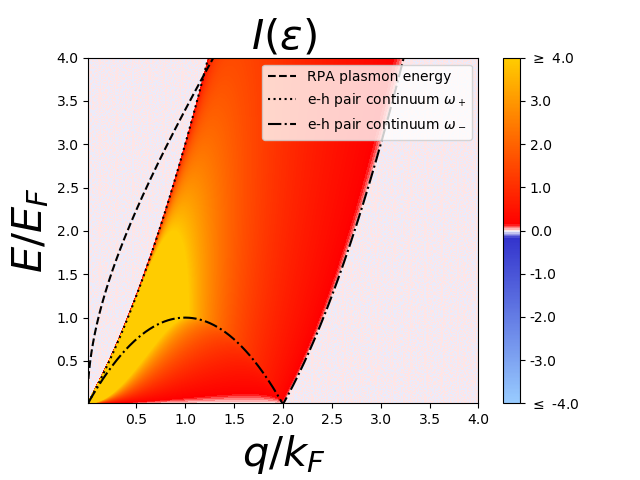}
\caption{2D RPA $(\vec{\bf q},\omega)$ for Ag}\label{fig4}
\end{figure}

\section{Summary}
The MsSpec DFM module is an independent module of the MsSpec program package \cite{a1}. It provides many methods to compute the dielectric function and related quantities 
for spectroscopies or other purposes. In particular, it 
allows a flexible way to calculate the fluctuation
potential that enters the quasi-boson modeling of the plasmon description and the plasmon field pioneered by
Hedin, Fujikawa and coworkers \cite{a2,a3}. But this module goes much beyond the particular problem of the embedding of plasmons within the multiple scattering framework, 
and can also be useful, for instance for classical liquids and plasmas. Many properties such as local field corrections, pair correlation functions, dielectric structure factors along with mean free paths, loss functions, electron-hole pair distributions etc, in addition to the dielectric functions can be computed from this module by selecting the corresponding switch in the input section of the input data file.

The present version is essentially restricted to 3D and 2D dielectric functions, but upgrades including new features will be made regularly. They will be available 
at the MsSpec program package website \cite{msspec}.   Other dimensionalities will be added through different choices of confinement. The main input parameter is the Wigner Seitz radius. Later addition will include electric and magnetic fields as well as temperature and effective mass. Among the various quantities that can enter the DFM, the present version computes LFC, structure factor, exchange correlation energy, e-e relaxation, pair correlation, memory, Nevanlinna function which has been published in the recent article~\cite{a36}.



\section{Acknowledgement}

A. M. is grateful to Rennes Métropole for providing her with two 6-month grants as a visiting scientist and generative scientific discussions during the period.









\bibliographystyle{elsarticle-num}



\end{document}